\documentclass[]{article}
\usepackage{spie}  

\input{psfig}

\def\ap{$''/$pix}
\def\micron{$\mu$m}
\def\ltsim{\:{_<\atop{^\sim}}\:}
\def\teff{${\rm T}_{\rm eff}$}
\def\deg{$^{\circ}$}

\title{Achieving a wide field near infrared camera for the Calar Alto 3.5m telescope}

\author{Coryn A.L.\ Bailer-Jones, Peter Bizenberger, Clemens Storz
\skiplinehalf
Max-Planck-Institut f\"ur Astronomie, 
K\"onigstuhl 17, D-69117 Heidelberg, Germany \\
}
\authorinfo{Send correspondence to Coryn Bailer-Jones, calj@mpia-hd.mpg.de\\
\hspace*{1.8em}A colour version of this paper is available from
http://www.mpia-hd.mpg.de/homes/calj/spie2000.html}

\begin{document}
\maketitle

\begin{abstract}
The ongoing development of large infrared array detectors has enabled
wide field, deep surveys to be undertaken.  There are, however, a
number of challenges in building an infrared instrument which has both
excellent optical quality and high sensitivity over a wide field. We
discuss these problems in the context of building a wide field imaging
camera for the 3.5m telescope at Calar Alto with the new 2K$\times$2K
HgCdTe HAWAII-2 focal plane array.  Our final design is a prime focus
camera with a 15$'$ field-of-view, called Omega 2000.  To achieve
excellent optical quality over the whole field, we have had to
dispense with the reimaging optics and cold Lyot stop.  We show that
creative baffling schemes, including the use of undersized baffles,
can compensate for the lost K band sensitivity. A moving baffle will
be employed in Omega 2000 to allow full transmission in the
non-thermal J and H bands.
\end{abstract}



\section{SCIENTIFIC MOTIVATION}

There are numerous scientific projects which would benefit from large
area infrared surveys. Most fit into the category of discovering and
characterising new objects. An example is a survey for very low mass
stars, brown dwarfs and free floating giant planets in open clusters
and star forming regions.  All of these objects are cool (\teff
$\ltsim 3000$) and have a significantly larger J than R or I band
flux. They can thus be detected with their optical--infrared colour,
or even their Z--J colours obtainable on the same HgCdTe detector.

Other areas of science which would benefit from surveys include: the
initial mass function of star forming regions; the dark matter content
and the age of the Galaxy from cool white dwarfs in the Galactic disk
and halo; Galactic structure traced via K and M giants (which, due to
the lower extinction in the near infrared, can be traced to larger
distances); galaxy surveys at low Galactic
latitudes; quasar surveys; star formation history and damped Lyman
alpha galaxies; high redshift galaxies; gravitational lenses and the
cosmological constant.

To serve these scientific goals, we plan to build a wide field near
infrared (0.8--2.5\micron) imaging camera for doing large area
surveys. Given the nature of the Calar Alto Observatory as a resource
for German and Spanish astronomers, this camera (Omega 2000) is
intended for use in common-user mode rather than undertaking
pre-defined surveys.

Some projects are less concerned with area coverage than with volume,
in which case deep `pencil-beam' surveys are more suitable. This may
be the appropriate strategy when searching for objects at a range of
distances, and in some cases may be more efficient than shallower wide
field surveys \cite{herbst_99a}. A list of current and future near
infrared survey facilities is given in Table~\ref{wfnircams}.

The rest of this paper is as follows. After giving the important
characteristics of the detector, we discuss the general issues
influencing the design of the instrument. Much attention is paid to
baffling schemes to minimise thermal radiation from the telescope
structures. The chosen optical design for Omega 2000 is then
presented, a long with a brief discussion of the electronic and
software systems and readout modes required for this high data rate
instrument.

\begin{table}
\begin{center}
\caption{A selection of current and future near infrared instruments.
All use HgCdTe arrays except NIRC and NIRI which use InSb arrays
(sensitive over 1--5\micron).  `Speed' is the product of field size
(in sq.\ degrees) and aperture (in m$^2$, neglecting the hole in the
primary mirror) and is some measure of how rapidly a survey area can
be covered to a certain depth.  It is at best a rough estimate as it
neglects instrument sensitivity and throughput and site
conditions. Some instruments have different cameras (and hence pixel
scales): the largest is shown.}
\label{wfnircams}
\vspace*{1ex}
\begin{tabular}{llllrrrl}
\hline
Instrument	& Telescope		& Aperture	& Detector	& Pixel scale	& Field size	& `speed' & First	 \\
		& and focus		& m		& 		& \ap\	& arcmin$^2$	&  		  & light	 \\
\hline
NIRC		& Keck I Cassegrain	& 9.8		& 1 256$\times$256	& 0.15  & 0.4           & 0.008		& operating     \\
NIRI		& Gemini North		& 8.1		& 1 1K$\times$1K	& 0.12	& 3.9		& 0.06		& mid 2000 	\\
SOFI		& NTT Nasmyth		& 3.6		& 1 1K$\times$1K	& 0.14  & 5.7		& 0.016		& operating     \\
ISAAC		& VLT UT1 Nasmyth	& 8.2		& 1 1K$\times$1K 	& 0.15	& 6		& 0.09		& operating     \\
Omega Cass	& CA 3.5m Cassegrain	& 3.5		& 1 1K$\times$1K 	& 0.30	& 26		& 0.07		& operating 	\\
Omega Prime	& CA 3.5m Prime		& 3.5		& 1 1K$\times$1K 	& 0.40	& 47		& 0.13		& operating	\\
IRIS2		& AAT Cassegrain	& 3.9		& 1 1K$\times$1K	& 0.45	& 59		& 0.20		& end 2000 	\\
CIRSI		& WHT Prime		& 4.2		& 4 1K$\times$1K 	& 0.32	& 119		& 0.46		& operating     \\
NIRMOS		& VLT UT3 Nasmyth	& 8.2		& 4 2K$\times$2K 	& 0.21	& 192		& 2.82		& end 2002	\\
Omega 2000	& CA 3.5m Prime		& 3.5		& 1 2K$\times$2K 	& 0.45	& 236		& 0.63		& mid 2001	\\
WFCAM		& UKIRT	Cassegrain	& 3.8		& 4 2K$\times$2K 	& 0.40	& 684		& 2.15		& end 2002	\\
VISTA		& VISTA	Cassegrain	& 3.9		& 9 2K$\times$2K 	& 0.30	& 936		& 3.19		& 2004    	\\
\hline
\end{tabular}
\end{center}
\end{table}

\section{INFRARED DETECTOR}

The detector is a major factor influencing the design of an infrared
camera, largely because of the limited choice and high cost of the
available arrays. The largest near infrared science grade array
currently available are 1K$\times$1K.  These are in use in a number of
existing cameras, such as the Omega Prime and Omega Cass cameras at
Calar Alto. Omega 2000 will use the next generation HAWAII-2
2K$\times$2K HgCdTe array from Rockwell~\cite{kozlowski_98a}. This is
very similar to the HAWAII array, except that it has slightly smaller
pixels (18.0\micron\ instead of 18.5\micron) and 32 rather than 4
outputs, allowing a faster readout (Table~\ref{hawaii2}). The arrays
are expected to have a very high filling factor (i.e.\ essentially no
`dead area' between the pixels).  Due to the background limited
observing conditions, the detector will be operated with a 1\,V reset
voltage to increase the full well capacity to about 200,000 electrons.

\begin{table}[h]
\begin{center}
\caption{Details of the HgCdTe photovoltaic HAWAII-2
focal plane arrays from Rockwell. Some values are the expected
performance based on the similar 1K HAWAII arrays.}
\label{hawaii2}
\vspace*{1ex}
\begin{tabular}{ll}
\hline
array size			& $2048 \times 2048$ pixels 	\\
pixel size			& 18\micron			\\
sensitivity range (QE$>$25\%)	& $\sim$0.8--2.58\micron	\\
mean QE over sensitivity range	& 50\%				\\
no.\ of outputs			& 32				\\
max.\ data rate			& $\simeq 1$\,MHz		\\
min.\ full frame read-out time	& $\simeq 0.13$\,s		\\
min.\ read noise		& $<5$\,e$^-$			\\
dark current (@ 77\,K)		& $<0.1$\,\,e$^-$/s		\\
full well capacity		& $\simeq$ 100,000\,e$^-$	\\
\hline
\end{tabular}
\end{center}
\end{table}

\section{DESIGN ISSUES}

\subsection{General Considerations}\label{considerations}

The primary science goal which Omega 2000 must address is wide field
imaging, and for a fixed size array this demands that we should have
the largest feasible pixel scale (acrseconds per pixel). ``Feasible''
in this case means (1) compatible with one of the focal stations
(prime or Cassegrain) on the 2.2m or 3.5m telescopes at Calar Alto;
(2) able to produce high quality images; (3) producible within an
acceptable timescale ($<$2 years) and budget ($\ltsim$1 million
DM). The timescale is set by the need to bring new technology into
prompt scientific use and the delivery time of the science grade array
(summer 2001).

Optical design problems aside, the upper limit on the pixel scale is
set by requiring a sufficient sampling of the PSF to permit accurate
photometry.  Too large a pixel scale means that the PSF is
undersampled, resulting in increased photometric errors.  A low
filling factor of the arrays would increase errors further.  The size
of the PSF is set by the seeing rather than diffraction at these
wavelengths and this aperture size. Only limited seeing statistics are
available for the 3.5m on Calar Alto: over the period 1993 to 1995,
the median near infrared seeing was 1.0$''$, and only better than
0.8$''$ 22\% of the time. (These values may be slightly optimistic as
they rely on integration times of less than 3s.) Sampling theory
specifies that at least two pixels should span the FWHM of the
PSF. However, in order to remove the variable background level common
to infrared imaging, it is usually necessary to take multiple images
of a given field with non-integer pixel offsets (dithers) between
them.  This enables a reconstruction of the PSF through sub-pixel
sampling, using methods such as those employed with the undersampled
WFPC2 camera on HST (e.g.\ the `drizzle'
algorithm~\cite{fruchter_97a}).  Thus pixel scales of between 0.4\ap\
and 0.5\ap\ were considered for Omega 2000 on the grounds that they
would only give a slight undersampling for a small fraction of the
time.

The financial constraints of this project require that the instrument
make use of existing telescope optics. Furthermore, as this will only
be one of several instruments on the telescope, modifications to the
telescope itself are not permitted.  The design of our system is
therefore limited to `conventional' prime or Cassegrain focus
solutions. If these constraints are relaxed, more sophisticated optical
designs which allow good optical quality across a large (c.\ 1 deg.)\
field-of-view become possible.  An example is a three-mirror telescope
combined with a Schmidt-type corrector plate, as will be used by the
wide field near infrared camera for UKIRT (WFCAM, see Atad-Ettedgui et
al., these proceedings) and the dedicated infrared/optical survey
telescope VISTA (Table~\ref{wfnircams}). Although the former has made
use of an existing telescope it requires a new f/9 secondary mirror,
and has some other drawbacks (see section~\ref{baffling}).

The dominant noise source for ground-based infrared imaging cameras is
usually photon noise from the background.  A significant part is from
the bright sky, which (at night) is predominantly OH airglow (below
3\micron)\ or thermal emission (above 3\micron)~\cite{papoular_83a}.
However, thermal radiation from the warm (0--15\deg\,C) telescope
mirrors, structure and dome is significant longwards of about
2.2\micron, i.e.\ for the K band~\cite{ramsay_92a}.  Therefore, as
much of the optics as possible should be enclosed in a cold
environment (a dewar). Often, the pupil is reimaged onto a cold Lyot
stop to minimise the amount of radiation reaching the detector from
structures around the mirror surfaces.  However, the complex optics
required for pupil reimaging can degrade the optical quality of the
image to unacceptable levels as the field of view and/or pixel scale
is increased.  Nonetheless, a wide field of view can be achieved by
dispensing with the reimaging optics and working at low f-ratio, for
example at prime focus, but at the expense of a larger background in
the K band.

There is, therefore, a trade-off between K band sensitivity,
field-of-view (or pixel scale) and optical quality. This trade-off can
be seen with the Omega Prime and Omega Cass infrared cameras at Calar
Alto (Table~\ref{wfnircams}). The former at prime focus has 1.8 times
the field of view of the latter at Cassegrain focus, but is less
sensitive in K.  We stress that this trade-off is only relevant for
the K band.  In the J and H bands, thermal radiation from the warm
surfaces is negligible compared to the OH emission from the sky.  Thus
if we only wanted a wide field imaging camera which operated up to
about 2\micron, prime focus would be the best option.

Weighing up these opposing factors is difficult, especially for a common-user
instrument which will be used for a whole range of science
projects. However, the scientific emphasis is on wide field imaging,
and provided the K band problem is not too severe (see
section~\ref{baffling}) a large field was felt to be a more
significant requirement.  Furthermore, a reimaging Cassegrain focus
solution would require more optics (minimum of 8 aspherical lenses)
and probably a non-standard dewar, increasing project time and cost by
about 50\%.

We also investigated a non-reimaging Cassegrain focus design. As with
the prime focus design, extra background from around the pupil can be
seen by the detector. If the secondary mirror is undersized then the
secondary is the pupil, and the extra background comes from the cold
sky. This is a much lower background flux than the warm floor/dome,
although the beam from the primary is vignetted by the undersized
secondary. If an exact-sized secondary mirror is used the beam is not
vignetted, but the camera would see some floor/dome via the
secondary. However, in both cases the reduction of the beam from f/10 (from
the existing secondary mirror) to f/2.35 (to achieve 0.45\ap)\ gave
extremely poor optical quality, so this design had to be rejected.

\subsection{Baffling a Non-reimaging Camera}\label{baffling}

The remaining design choices for Omega 2000 are therefore a
Cassegrain focus camera with a cold Lyot stop and a no-cold stop prime
focus camera. In this section we investigate the relative sensitivities
and survey speeds of such cameras and as well as different baffling
schemes for prime focus cameras.

Each detector pixel receives all the light from the $2\pi$\ steradians
(hemisphere) in the direction of the dewar window. Most of this is
cold, dark dewar which contributes negligible radiation. With a Lyot
stop, the rest of the light is from the pupil, but when it is omitted,
radiation from the warm surfaces (telescope structure, floor/dome
etc.)\ around the pupil reaches the detector, and the noise in this
radiation lowers the instrument's K band sensitivity.

\begin{figure}[t]
\hbox{\hspace{0.25\textwidth}
\psfig{figure=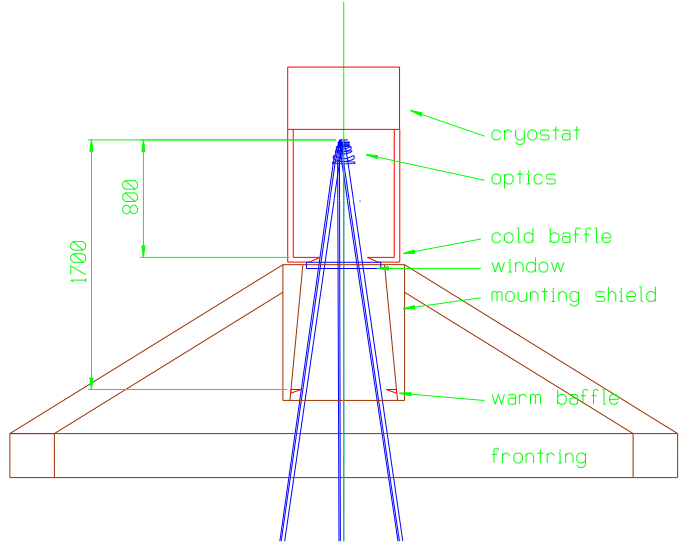,width=0.50\textwidth,angle=270}
}
\caption{Prime focus camera with cold and warm baffles}
\label{bafflepic}
\end{figure}

This radiation can be reduced in a prime focus design by placing a
cold annular baffle between the detector and primary mirror
(Fig.~\ref{bafflepic}).  The further this baffle is from the detector
(and the nearer it is to the pupil), the smaller is the solid angle
subtended by warm surfaces at the detector. The ideal place for the
baffle is at the primary mirror, but a cold baffle of 3.5m radius is
not feasible!  The only practical place is therefore inside the
detector dewar, and its maximum distance from the detector is set by
one or more of the following:
\begin{enumerate}
\item{not vignetting the beam reaching the primary mirror from the
sky,}
\item{the optical quality, mechanical stability and
availability of the large dewar window,}
\item{the cost of and heat load on the increasingly larger dewar.}
\end{enumerate}

\begin{figure}[t]
\hbox{\hspace{0.25\textwidth}
\psfig{figure=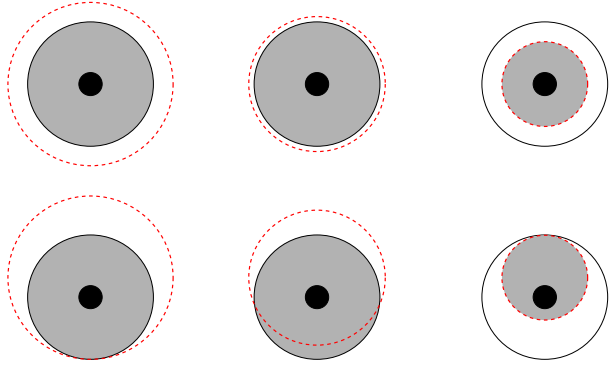,width=0.45\textwidth,angle=0}
}
\caption{Explanation of baffling schemes.  The primary mirror plane as
seen from the center of the detector (top row) and corner of the
detector (bottom row). The solid black circle and small filled black
circle are the primary mirror and its central hole respectively, and
are of course the same size in all cases.  The dashed (red) circle is
the projection of a baffle onto the primary mirror plane for the three
cases of critical vignetting (left), intermediate vignetting (middle)
and super vignetting (right). The grey area is the area of the primary
mirror seen by the detector; the area between the solid (black) and
dashed (red) lines is either floor or baffle, depending on the
vignetting mode.
}
\label{vigmodes}
\end{figure}

Typically one would use a baffle which is just large enough so as not
to obstruct the view of the primary mirror from any point on
the detector.  This condition we call {\em critical vignetting}.  Note
that the larger the field-of-view, the larger this baffle must be in
order to not vignet, and so the larger the solid angle of warm
floor/dome which can be seen.  This puts a sensitivity penalty on
large pixel scales at prime focus.  As the baffle is decreased in
radius from its critically vignetting size, the floor/dome solid angle
is reduced but at the expense of vignetting the light from the primary
mirror (Fig.~\ref{vigmodes}). Note that the amount of primary mirror
which can be seen by a given pixel depends on the location of that
pixel on the detector.  This gives rise to differential vignetting
across the field of view.  As the baffle is made smaller still, there
comes a point at which no floor/dome can be seen from any point on the
detector and there is nothing gained in reducing the baffle radius
further. This condition we refer to as {\em super vignetting}, and has
the advantage of uniform vignetting across the whole field of view.

\begin{figure}[t]
\hbox{
\psfig{figure=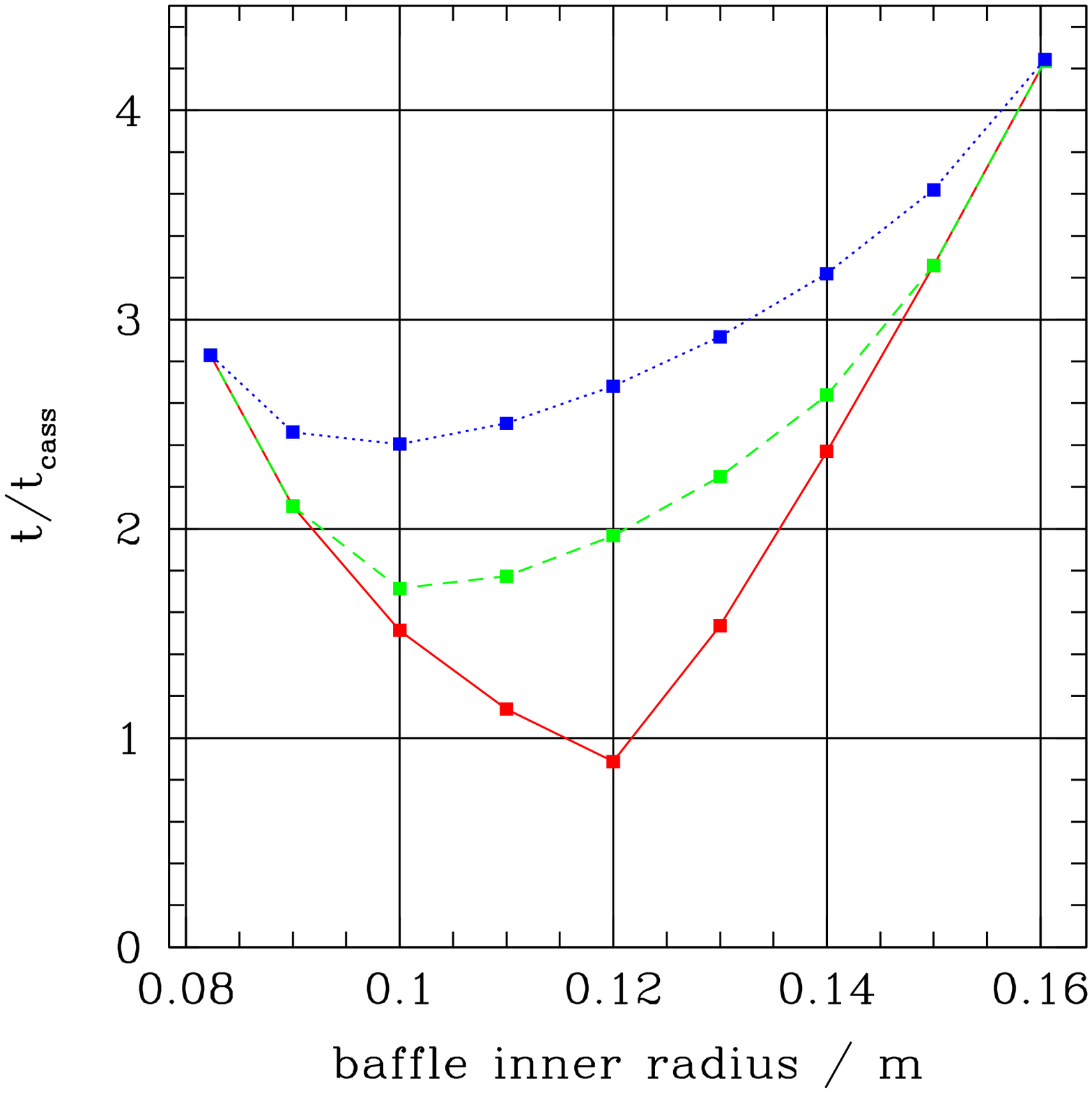,width=5.6cm,angle=0}
\hspace{0.2cm}
\psfig{figure=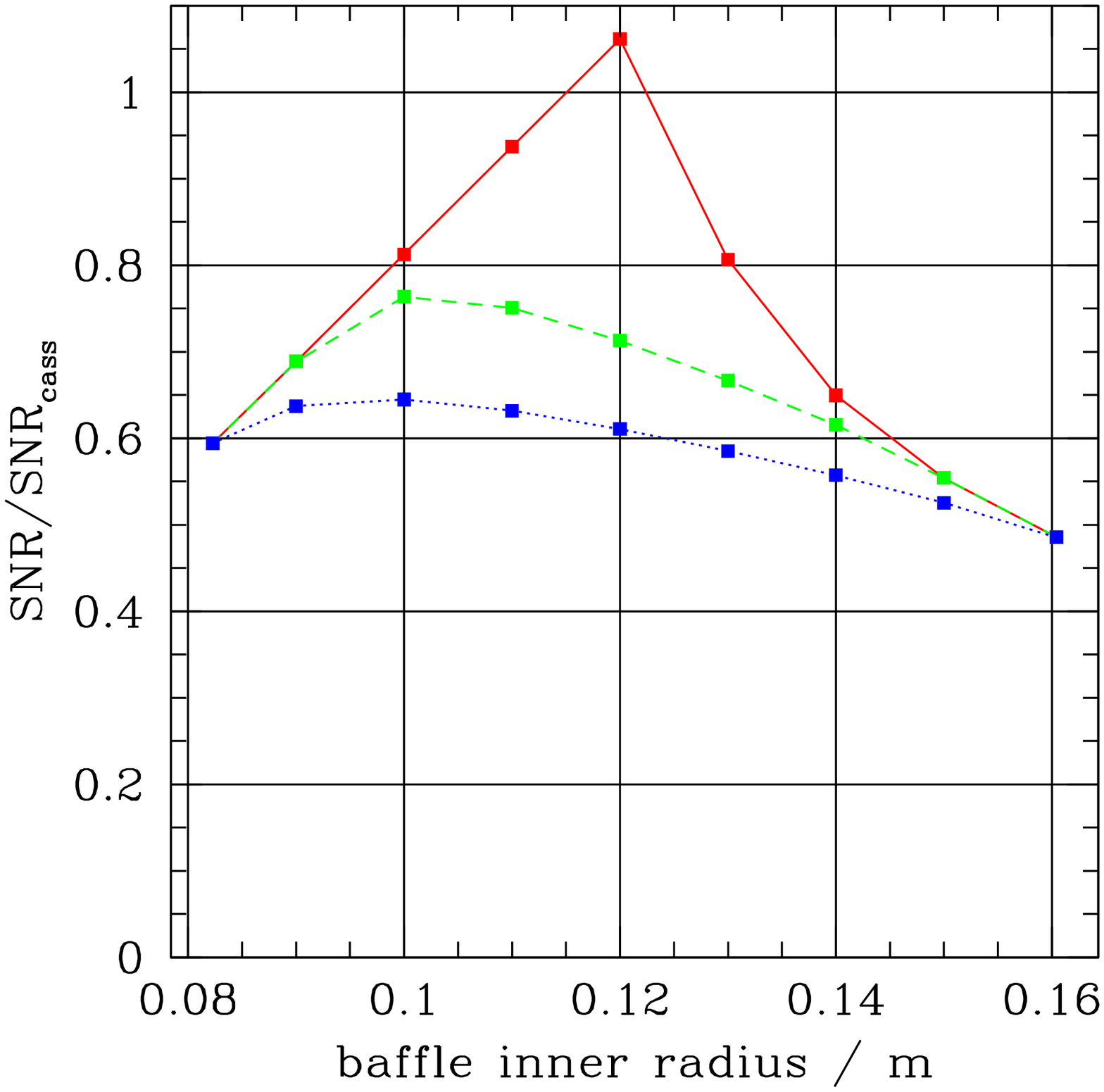,width=5.6cm,angle=0}
\hspace{0.2cm}
\psfig{figure=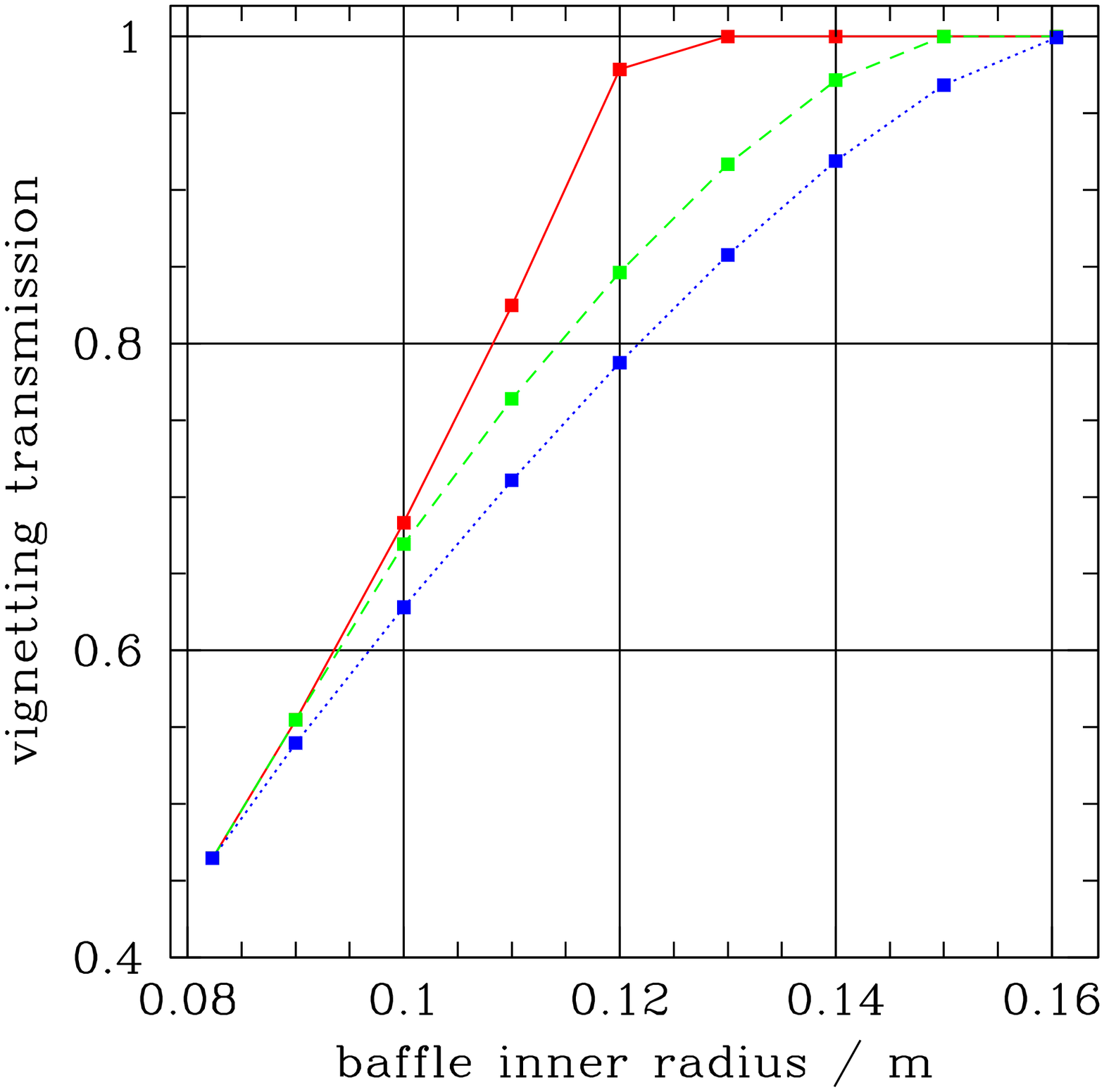,width=5.6cm,angle=0}
}
\caption{K band sensitivities of a 0.45\ap\ prime focus camera
relative to a perfectly baffled Cassegrain focus camera with the same
pixel scale.  The prime focus camera has a cold baffle 0.8m from
detector with inner radius shown on the horizontal axes.  $t/t_{\rm
cass}$ is the relative integration time to reach a given
signal-to-noise ratio (SNR), and ${\rm SNR}/{\rm SNR}_{\rm cass}$ is
the relative SNR in a given integration time, both for a background
limited source. Each plot shows the characteristics for different
points on the 2K$\times$2K detector: center (red solid); edge (green
dashed); corner (blue dotted). The vignetting transmission fraction is
the fraction of source light from the primary which reaches the
detector. The critical and super vignetting points are at radii of
0.160m and 0.082m respectively.}
\label{varycoldbaf}
\end{figure}

We have done some simulations to see how the sensitivity of an
instrument on the 3.5m varies between these two extreme vignetting
conditions.  With a 0.45\ap\ prime focus camera and a cold baffle
fixed 0.8m from the detector, its inner radius is varied between 160mm
(critical vignetting) and 82mm (super vignetting). The calculation
assumes that the background is due to thermal radiation from the
primary mirror (r=1.75m, $\epsilon$=0.06) and its central hole
(r=0.325m, $\epsilon$=1.0), and the signal is proportional to the area
of the primary which can be seen (minus its hole). Sources are assumed
to be black bodies at ambient temperatures.  The cold baffle is in the
dewar so contributes negligible radiation.  Radiation from the primary
support structure has also been neglected.  We further assume that OH
emission from the sky contributes an equal\footnote{This is what is
expected in summer. In winter the thermal radiation from the optical
surfaces in a perfectly baffled camera only accounts for about 20\% of
the total background, the rest being sky OH emission. Thus in winter
the performance of the prime focus camera relative to a Cassegrain
focus one will be better than that shown here.}  amount of background
radiation as the thermal emission from the primary mirror (McCaughran
et al., unpublished). Thermal emission from the sky at these
wavelengths is negligible compared to the OH
emission. Fig.~\ref{varycoldbaf} shows how the sensitivity of the
instrument relative to a perfectly baffled Cassegrain focus camera
changes with baffle size.  This perfect camera has only thermal
emission from the primary mirror and hole, plus thermal emission 
from an exact-sized secondary mirror with an assumed emissivity of
0.03. As a reimaging Cassegrain focus camera would have several more
lenses, it would have a slightly lower throughput, so the
sensitivities of the prime focus systems relative to this are slightly
conservative.

These simulations show that the prime focus camera is almost always
less sensitive than a perfectly baffled Cassegrain focus camera in the
K band. Interestingly, however, the performance of the prime focus
camera is improved (sometimes considerably) by using an undersized
baffle. The optimum baffling radius is different for different points
on the detector, because the projection of the baffle onto the primary
mirror is different for each detector point.

\begin{figure}[h]
\vspace*{0.5cm}
\hbox{
\psfig{figure=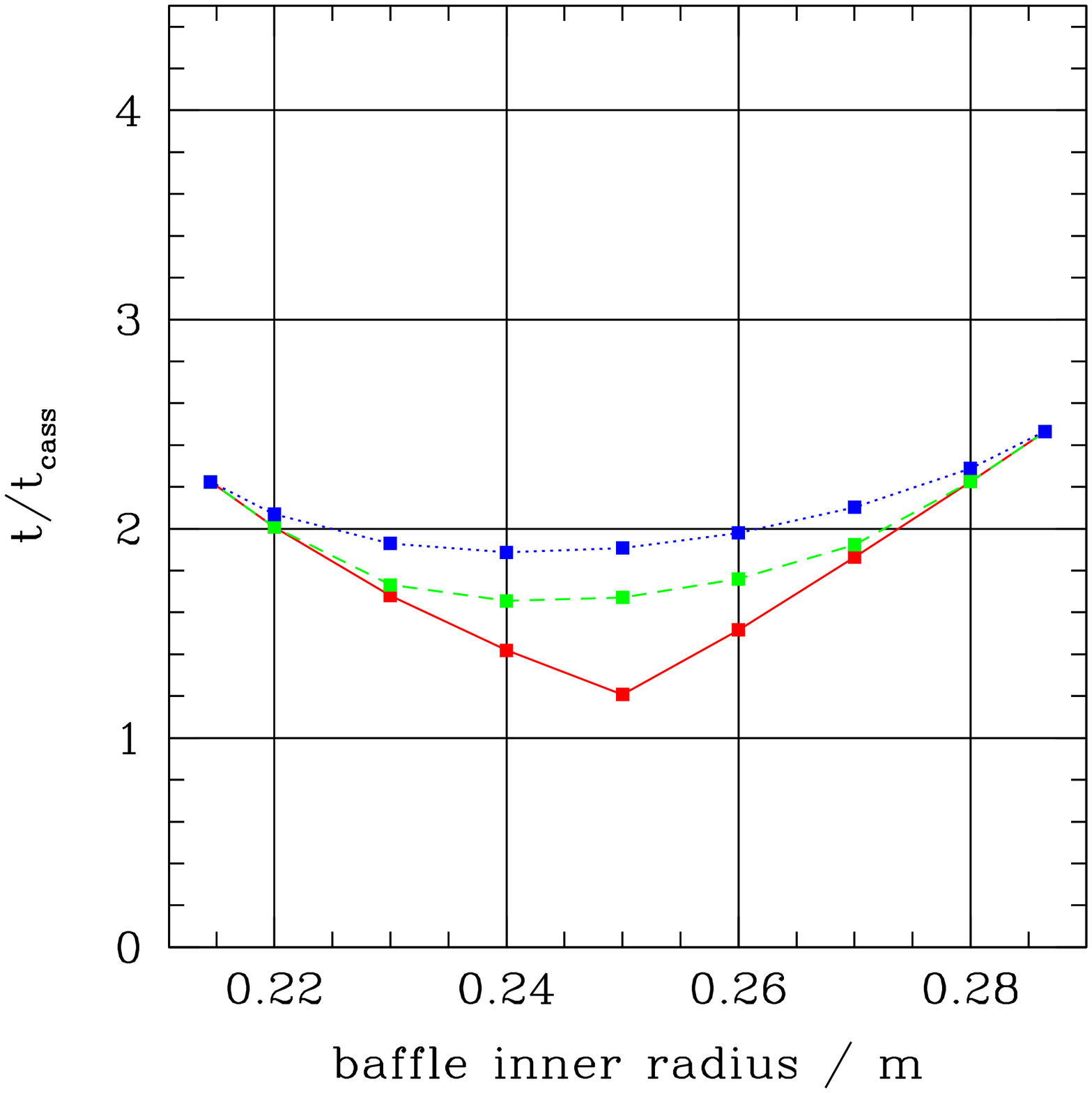,width=5.6cm,angle=0}
\hspace{0.2cm}
\psfig{figure=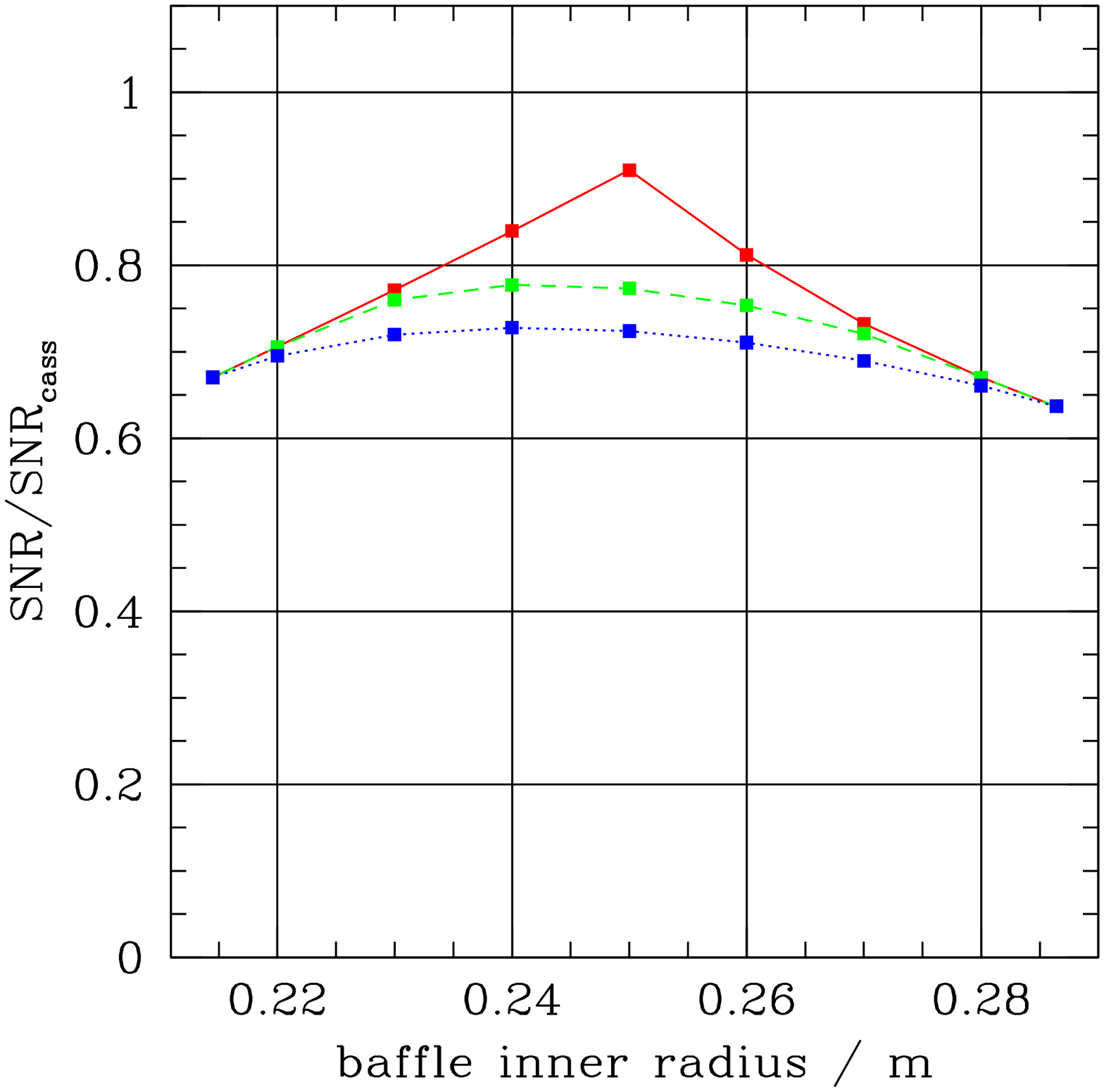,width=5.6cm,angle=0}
\hspace{0.2cm}
\psfig{figure=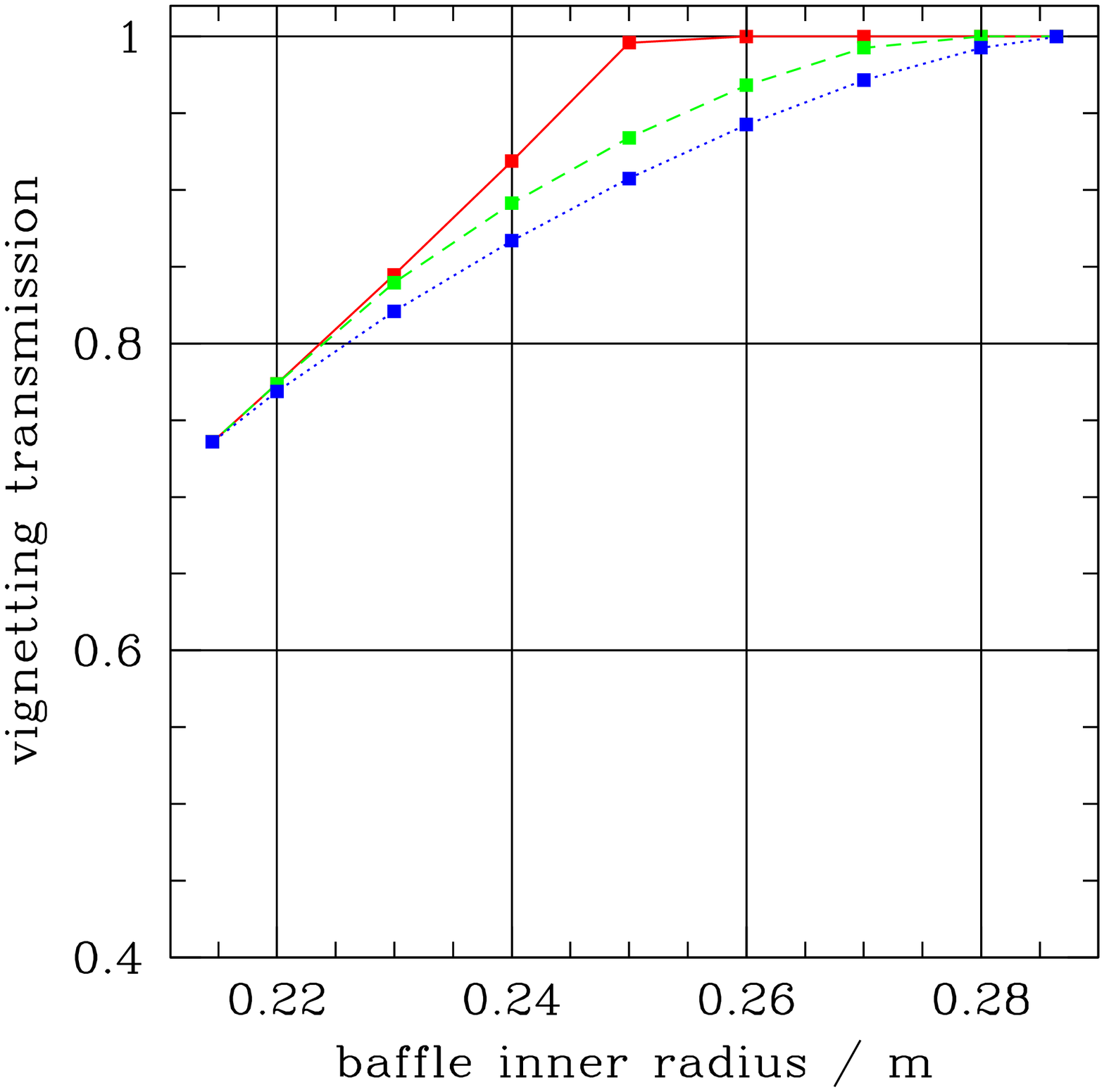,width=5.6cm,angle=0}
}
\caption{Same as Fig.~\ref{varycoldbaf} but now a critically
vignetting cold baffle is fixed 0.8m from the detector, and a warm
baffle (emissivity $\epsilon$=0.10) 1.7m from the detector is varied
in radius between its super (r=0.215) and critical (r=0.286) vignetting points.}
\label{varywarmbaf}
\end{figure}

The situation can be improved by adding a warm baffle further away
from the detector. The ideal case would be to put a 3.5m diameter high
reflectivity ring around the primary mirror which looks through the
dome slit at the cold sky. But as it would need an outer radius of
almost 4.6m, it would be very expensive and highly impractical.  Hence
the warm baffle must hang as far below the dewar window as is possible
without vignetting the beam from the sky, which is about 1.7m for the
Calar Alto 3.5m (see section~\ref{optical_design}).  As in Omega
Prime, the warm baffle is a polished ellipsoidal annulus which looks
into the dewar window to avoid collecting radiation from warm
surfaces~\cite{bizenberger_98a}.  Gold coating will further lower its
emissivity, but we conservatively assume $\epsilon$=0.10
here. Fig.~\ref{varywarmbaf} shows how the sensitivity varies as we
change the size of this baffle. Note that a critically vignetting cold
baffle is also included to reduce the amount of the warm baffle seen
by the detector. Because the warm baffle is further from the detector,
the variation in sensitivity both for different pixel positions and
baffle sizes is less than in Fig.~\ref{varycoldbaf}.  Moreover, the
sensitivity relative to the perfect Cassegrain focus camera is better.
Uniform sensitivity across the whole field-of-view is usually
desirable for a survey instrument, so in practice the super vignetting
case may be preferred to intermediate vignetting.  

\begin{table}[t]
\begin{center}
\caption{K band sensitivities for different baffling schemes for a
prime focus camera without a Lyot stop.  A 2K$\times$2K detector with
18\micron\ pixels is assumed.  Each set-up has a warm baffle fixed
1.7m from the detector and a cold baffle (inside the dewar) either
0.8m or 1.0m from the detector. The ``vignetting mode'' is explained
in the text: for example, ``warm, super'' means that the warm baffle
super vignets the detector. The vignetting transmission is the fraction
of source light from the primary which reaches the detector.
t/t$_{\rm cass}$ is the time required to reach a fixed SNR in the background
limit relative to a perfectly baffled Cassegrain focus
camera. SNR/SNR$_{\rm cass}$ is the SNR achieved in a fixed
integration time in the background limit relative to a perfectly
baffled Cassegrain focus camera.}
\label{baffletab}
\vspace*{1ex}
\begin{tabular}{lllllll}\cline{3-4}
\hline
pixel scale	& cold baffle	& \multicolumn{2}{l}{vignetting mode}& vignetting	& t/t$_{\rm cass}$ 	&	SNR/SNR$_{\rm cass}$ \\
	 	& distance	& 			&		& transmission	&			&				\\
\hline
0.45\ap	& 0.8m		& cold & critical	& 1.00		& 4.24		& 0.49			\\
0.45\ap	& 0.8m		& cold & super		& 0.46		& 2.84		& 0.59			\\
0.45\ap	& 0.8m		& warm & critical	& 1.00 		& 2.46		& 0.64			\\
0.45\ap	& 0.8m		& warm & super		& 0.74		& 2.23		& 0.67			\\
0.45\ap	& 1.0m		& cold & critical	& 1.00		& 3.48		& 0.54			\\
0.45\ap	& 1.0m		& cold & super		& 0.56		& 2.09		& 0.69			\\
0.45\ap	& 1.0m		& warm & critical	& 1.00		& 2.39		& 0.65			\\
0.45\ap	& 1.0m		& warm & super		& 0.74		& 2.09		& 0.69			\\
\hline
0.40\ap	& 0.8m		& cold & critical	& 1.00		& 3.63		& 0.52			\\
0.40\ap	& 0.8m		& cold & super		& 0.54		& 2.22		& 0.67			\\
0.40\ap	& 0.8m		& warm & critical	& 1.00		& 2.18		& 0.68			\\
0.40\ap	& 0.8m		& warm & super		& 0.78		& 1.89		& 0.73			\\
0.40\ap	& 1.0m		& cold & critical	& 1.00		& 3.01		& 0.58			\\
0.40\ap	& 1.0m		& cold & super		& 0.62		& 1.76		& 0.75			\\
0.40\ap	& 1.0m		& warm & critical	& 1.00		& 2.12		& 0.69			\\
0.40\ap	& 1.0m		& warm & super		& 0.78		& 1.79		& 0.75			\\
\hline
0.40\ap	& 0.4m		& warm$^*$ & critical	& 1.00		& 2.24		& 0.67			\\
0.40\ap	& 0.4m		& warm$^*$ & super	& 0.76		& 1.93		& 0.72			\\
\hline
\end{tabular}
\end{center}
\begin{center}
\begin{minipage}{13.5cm}
$^*$ These two cases have been calculated for Omega Prime, which has a
1K detector\\
\hspace*{1em}with 18.5\micron\ pixels and a warm baffle 0.82m from the detector.
\end{minipage}
\end{center}
\end{table}

As mentioned earlier, a larger field of view requires a larger baffle
for any given vignetting mode, and therefore a reduced camera
sensitivity.  To get a quantitative idea of this, we repeated the
above calculations for the super and critical vignetting modes but for
two different pixel scales, 0.40\ap\ and 0.45\ap, and with baffles at
various distances.  The results are shown in Table~\ref{baffletab} and
indicate that for a given pixel scale and baffle distances, super
vignetting provides better sensitivity than critical vignetting.
Having the cold baffle 1.0m rather than 0.8m from the detector gives
only a relatively small improvement, and the extra demands this
places on the system (e.g.\ larger dewar window) are probably not worth
it.  When super vignetting with the warm baffle, the 0.40\ap\ camera
is 1.18 times faster for a single shot than the 0.45\ap\
camera. However the 0.45\ap\ scale covers a large area 1.27 times
faster, with the net result that the 0.45\ap\ scale is slightly faster
for K band surveys. Of course, in the J and H bands super vignetting
should never be used, so implementation of super vignetting requires a
baffle which can be rapidly (and automatically) moved. Note that the
0.45\ap\ camera in super vignetting mode should have the same K band
sensitivity as the current Omega Prime camera (which has a critically
vignetting warm baffle). Interestingly, Omega Prime camera should be
about 15\% faster if it had a super vignetting warm baffle.  Such a
baffle has been made, but attempts to test it on the telescope have so
far been foiled by bad weather.

The 0.45\ap\ camera is still 2.2 times slower than a perfectly baffled
Cassegrain focus camera. However, as a pixel scale of only about
0.35\ap\ with good optical quality could be achieved at Cassegrain
focus, the survey speed of the perfect Cassegrain focus camera is only
about 1.3 times faster in K, and 1.7 times slower in J and H.  It
should be emphasised that these figures are rather sensitive to the
assumptions laid out above, in particular with regard to the
emissivity of the optical surfaces and the fraction of OH emission
from the sky. Lowering the emissivity or temperature of the warm
baffle in particular leads to much better sensitivity.

A three-mirror Cassegrain focus camera similar to that which will be
used by WFCAM on UKIRT (3.8m f/2.5 primary) would be possible for the
3.5m at Calar Alto. However, although the current WFCAM design has a
cold Lyot stop, it has a vignetting transmission of 73\% in all
wavebands (WFCAM web pages, ATC, Edinburgh).  This is very similar to
the K band transmission of Omega 2000 when using the super vignetting
warm baffle. Additionally, WFCAM has a center-to-corner distortion of
0.5\%, or 42 pixels. However, the super vignetting Omega 2000 design
does not appear to be extendible to the much larger field-of-view
obtainable with WFCAM.

\section{OMEGA 2000 DESIGN}

\subsection{Optical Design}\label{optical_design}

\begin{figure}[t]
\vspace*{1.5cm}
\hbox{\hspace{0.33\textwidth}
\psfig{figure=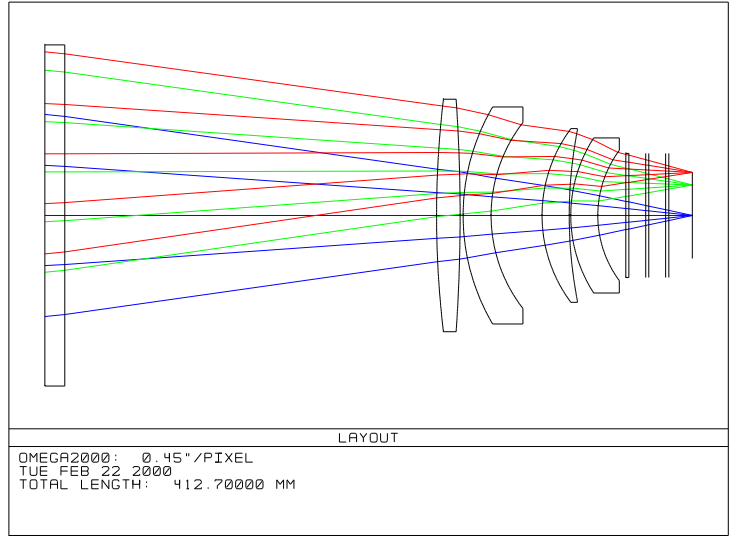,angle=90,width=0.45\textwidth} 
}
\caption{Ray trace diagram. The optical surface to the left is the dewar window,
followed by the four lenses, three filter wheels and finally the detector.
At any one time only one filter is in the optical path.}
\label{raytrace}
\end{figure}
\begin{figure}[t]
\vspace*{1.5cm}
\hbox{\hspace{0.33\textwidth}
\psfig{figure=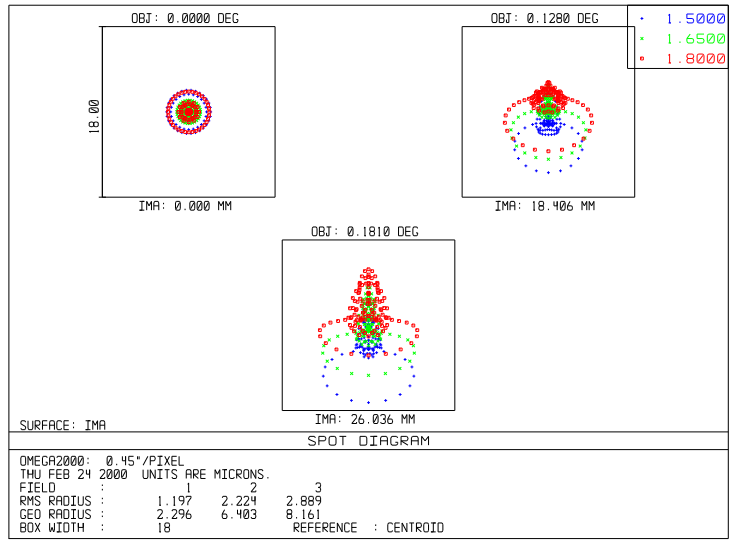,width=0.45\textwidth,angle=90}
}
\caption{Spot diagrams. Geometrically-traced points on a single pixel for rays from a point
source, traced through different parts of the optical system. The tracing has been done for
three monochromatic wavelengths in the H band, and for pixels in the center (0.0 deg), edge (0.128 deg)
and corner (0.181 deg) of the detector array.}
\label{spots}
\end{figure}
\begin{figure}[t]
\vspace*{1.6cm}
\hbox{\hspace{0.35\textwidth}
\psfig{figure=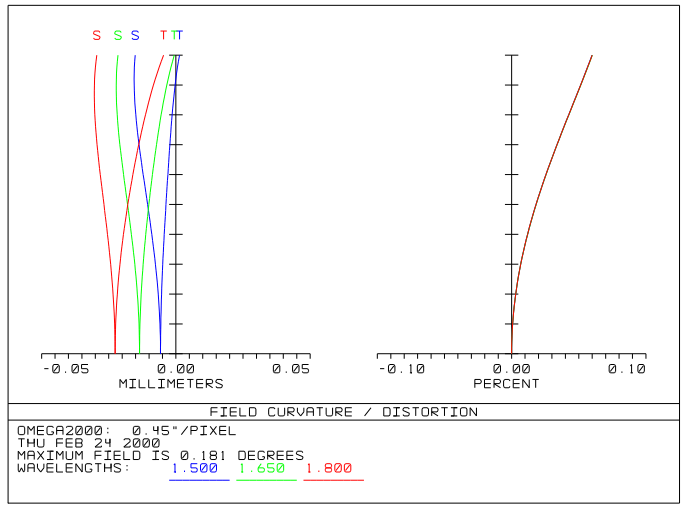,width=0.42\textwidth,angle=90}
}
\caption{Field curvature (left) and distortion (right) for three wavelengths in the H band.
The vertical axis is increasing distance from the center of the field (detector) to the corner.}
\label{distortion}
\end{figure}
\begin{figure}
\vspace*{0.7cm}
\hbox{\hspace{0.35\textwidth}
\psfig{figure=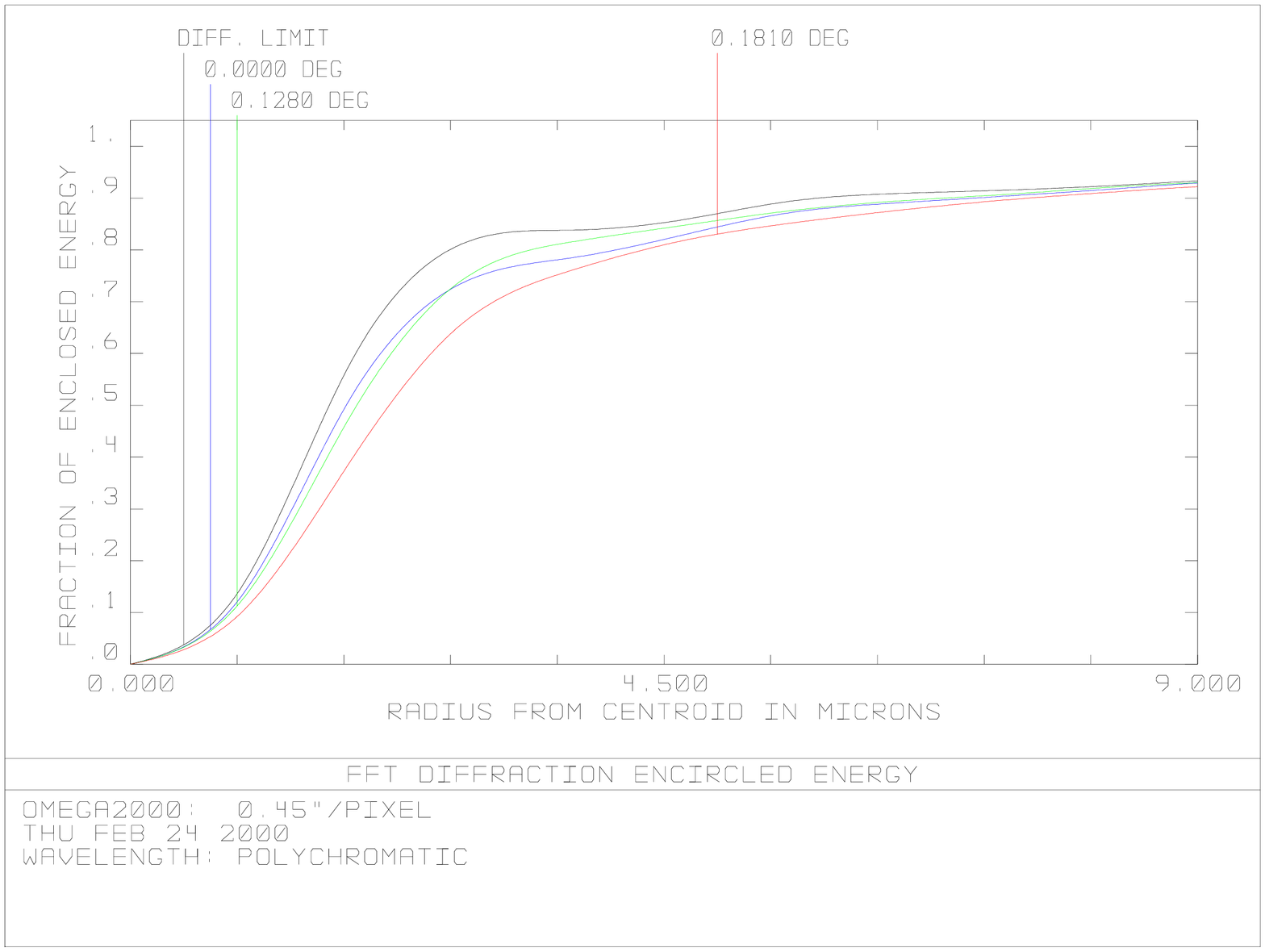,width=0.41\textwidth,angle=90}
}
\vspace*{8ex}
\hbox{\hspace{0.35\textwidth}
\psfig{figure=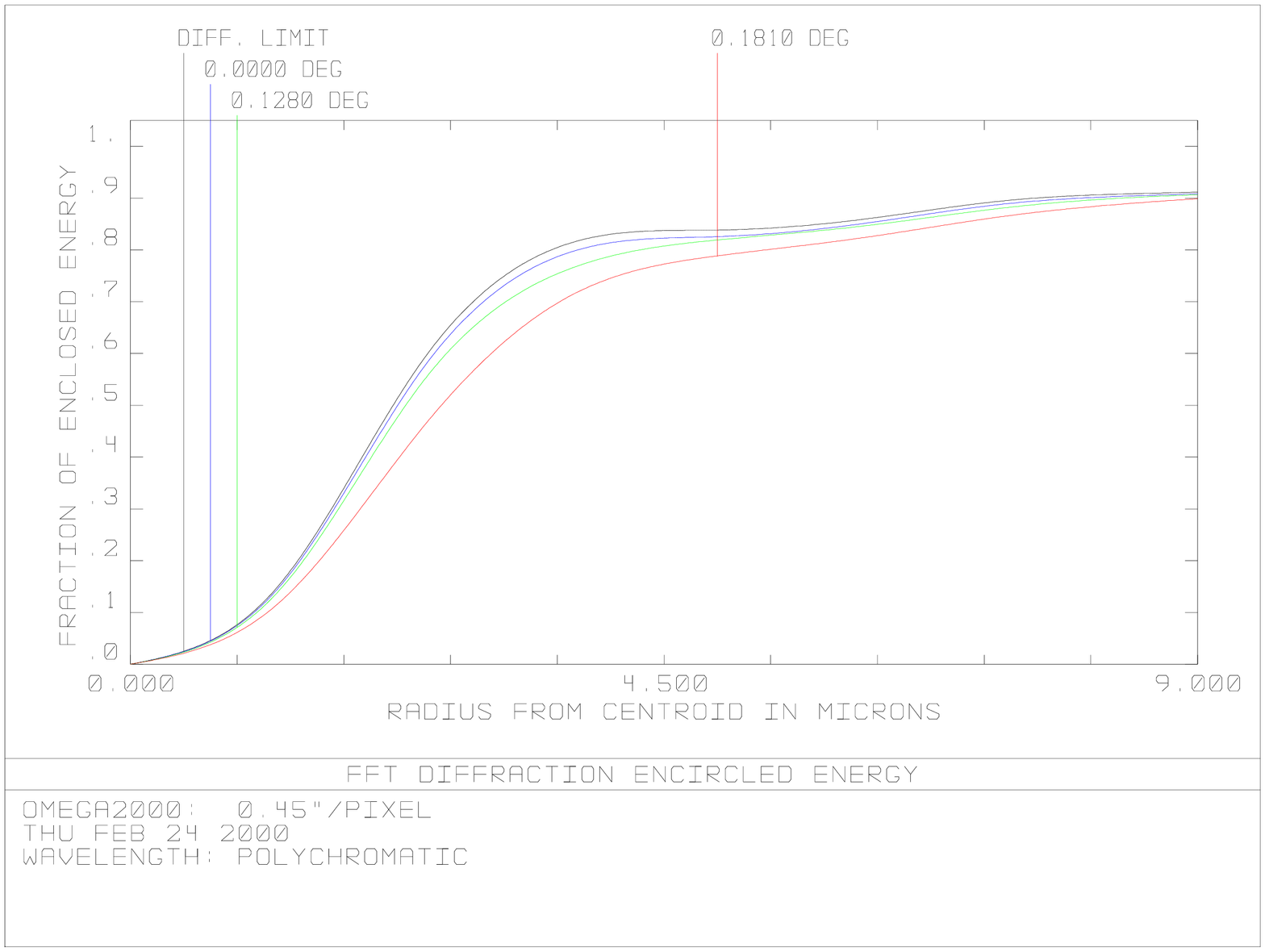,width=0.41\textwidth,angle=90}
}
\vspace*{8ex}
\hbox{\hspace{0.35\textwidth}
\psfig{figure=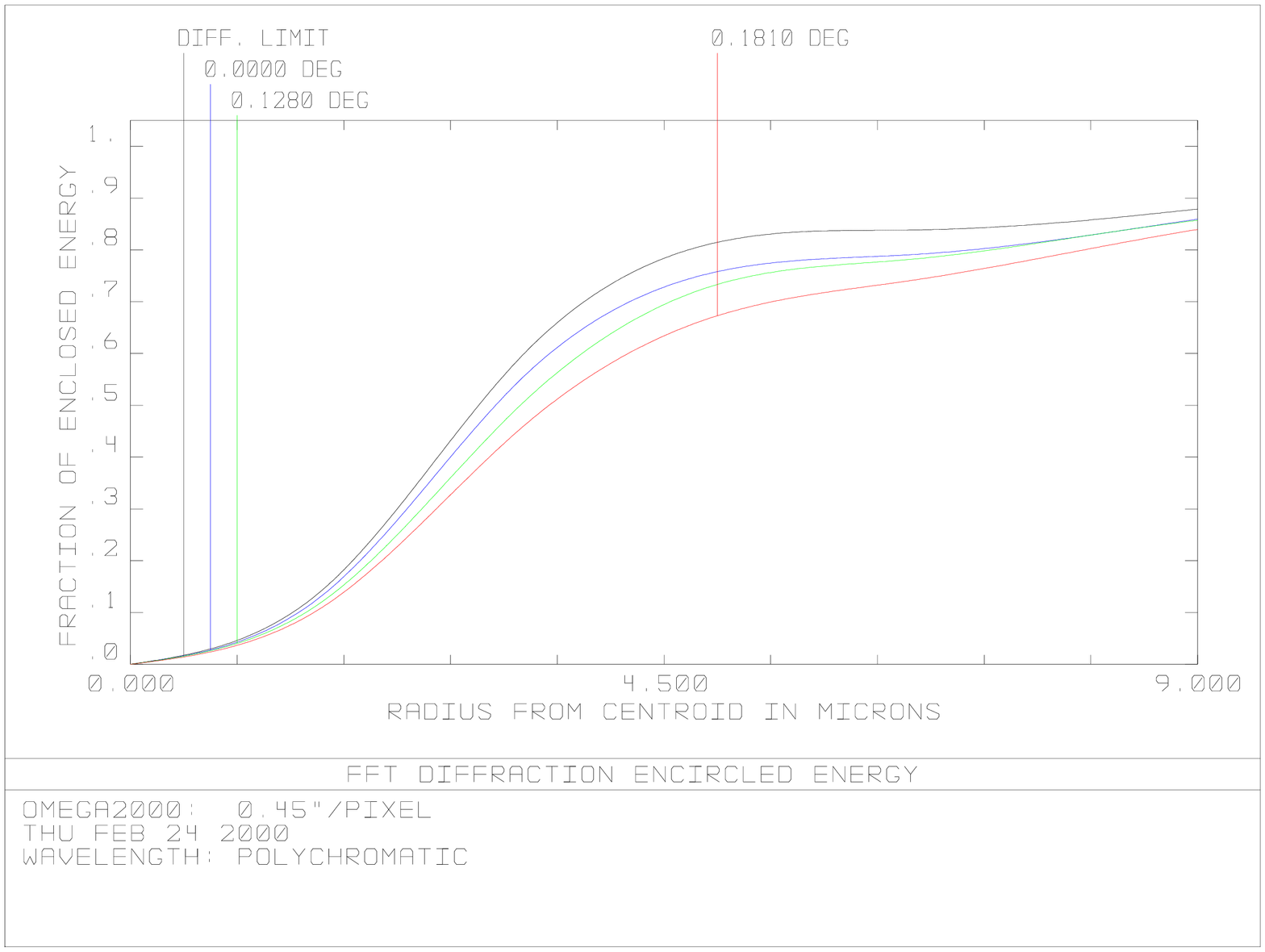,width=0.41\textwidth,angle=90}
}
\caption{Enclosed energy diagrams for the J band (top), H band
(middle) and K band (bottom).  These show the fraction of energy
enclosed within a given distance from the center of the image of a
point source.  The center-to-edge distance of a pixel is
9.0\micron. The top (black) line is the diffraction limit. The other
three lines (top to bottom) are for positions at the center, edge and
corner of the detector.}
\label{enclosed_energy}
\end{figure}

Following all of the considerations described in
section~\ref{considerations}, we decided to place Omega 2000 at the
prime focus of the 3.5m telescope with a 0.45\ap\ scale.  The optical
system consists of four lenses (Fig.~\ref{raytrace}) and has excellent
optical quality (Figs.~\ref{spots}, \ref{distortion}
and~\ref{enclosed_energy}). Three filter wheels are envisaged, each
with slots for six 3 inch (7.6cm) diameter filters.  The parameters for
the system are given in Table~\ref{params}.

\begin{table}
\begin{center}
\caption{Omega 2000 instrument parameters. The inner radii of the
baffles are for critical vignetting.}
\label{params}
\vspace*{1ex}
\begin{tabular}{ll}
\hline
primary mirror diameter 	&	3.5m				\\
primary mirror hole diameter 	&	0.65m				\\
primary mirror focal ratio	&       3.5				\\
focal station			& 	prime 				\\
no.\ of lenses			&	4 (cold)			\\
no.\ of optical surfaces	&	13 (2 warm)			\\
pixel scale			&  	0.45\ap 			\\
final focal ratio		&	2.35				\\
field size      		&	$15.4' \times 15.4'$ (236 sq.\ arcmin)	\\
dewar window radius		&	175mm				\\
cold baffle--detector distance	&	0.8m				\\
cold baffle inner radius	&	160mm				\\
warm baffle--detector distance	&	1.7m				\\
warm baffle inner radius	&	286mm				\\
\hline
\end{tabular}
\end{center}
\end{table}

The very low distortion of $<$0.06\% is less than 1 pixel
center-to-corner, permitting images to be overlaid with a simple x,y
shift. As demonstrated in the previous section, the sensitivity will
be very similar to Omega Prime: with a two minute integration in 1$''$
seeing, the central pixel of a point source is 5$\sigma$ above the
background noise at limits of J=19.2, H=18.1 and K'=17.5
magnitudes~\cite{bizenberger_98a}. This compares with a
mean sky brightness of {\em approximately} J=15.2, H=13.6 and
K'=13.0 (MAGIC instrument web pages, Calar Alto).

\subsection{Mechanical Design}

All of the optics will be enclosed in a dewar cooled with liquid
nitrogen. The design will be similar to that of Omega
Prime~\cite{bizenberger_98a} with two nested tanks to ensure a stable
temperature of 77\,K. The optimal baffling was discussed in
section~\ref{baffling}.
A fixed critically vignetting cold baffle is on the inside of the
dewar window. The window has a radius of 350mm. The cylindrical dewar
will be about 0.6m in diameter and between 1.5m and 2.0m long: the
exact figure will depend on the final cryogenic design.  The heat load
on the dewar is large, approximately 300W.  A second warm baffle
shaped as an oblate ellipsoid sits 1.1m from the dewar window
(i.e. 1.7m from the detector).  This distance corresponds with the
bottom of the existing light shield attached to the bottom of the
front ring.
This shield actually slightly vignets the beam from the sky, but as
it is required for other instruments it cannot be removed.  Although
the exact design has not yet been worked out, the warm baffle will be
moveable between the critical and super vignetting modes (or may
consist of two separate baffles).  The former mode will be used for
the J and H bands, and the latter for the K band. The projection of
the super vignetting baffle onto the primary mirror (as seen from the
detector) has a diameter of 3.0m, so does not threaten to diffraction
limit the instrument at the longest wavelength (2.5\micron).

\subsection{Electronics and Readout Modes}

A general readout electronic system is being developed in house.  This
will be used for a number of different instruments currently under
development at MPIA, including Omega 2000 and the MIDI interferometer
for the VLT. Only small modifications are required for different
detector types.

The HAWAII-2 detector works either with 1 signal channels per quadrant
or with 8 signal channels per quadrant simultaneously. All four
quadrants can be operated in parallel. The pixels can be clocked at up
to 1 MHz which will deliver 1 Megapixels per second per signal
channel, with each pixel represented as 16 bits.  The minimum
integration time is given by the frame readout time, which is the
normal state for arrays with integrate-while-read working mode. When
working with all 32 channels, the data rate will be 64 Mbyte/sec and
the full frame rate is about 8Hz. With only 4 channels (1 per
quadrant) the data rate is 4 Mbyte/sec and the full frame rate is
about 1Hz.

\begin{figure}[t]
\hbox{\hspace{0.23\textwidth}
\psfig{figure=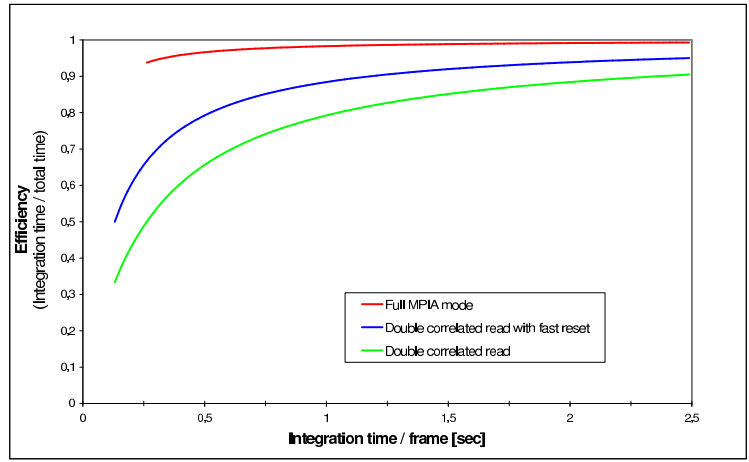,width=0.46\textwidth,angle=270}
}
\vspace*{1cm}
\caption{Comparison of the efficiency of the three read out modes as a
function of integration time for a stack of 15 frames for: full MPIA
mode (top), double correlated read with fast reset (middle) and
conventional double correlated read (bottom).}
\label{readouteff}
\end{figure}

The background limit will be reached in broadband imaging with the
HAWAII-2 array on Omega 2000 in a few seconds.  Many images
(``repeats'') are therefore required to achieve sufficient SNR, so it
is very important that the array can be read out quickly
with minimum dead time. It is also not necessary to use readout modes
which greatly reduce the read noise. There are three different double
correlated readout (DCR) modes suitable for detectors with non-destructive
readout modes, such as the HAWAII detectors, and each give very
different readout efficiencies:
\begin{enumerate}
\item{{\bf Double correlated read.}  This is the most conservative
mode, in that it passes throught the whole array three times to
achieve a single image. The first time the whole array is reset, in
the second the whole array is read to determine the pixel bias
(offset) values, and finally the third read measures the integrated
(light-exposed) values. The difference between the integrated and bias
values for each pixel is the signal. At minimum integration time the
efficiency (as measured by the ratio of integration time to total
cycle time required to achieve an image) is 33\%. To achieve an
integration time longer than the minumum, there is a pause after the
second reading of the whole array.}
\item{{\bf Double correlated read with fast reset.} This is the current
standard readout mode for MPIA instruments. In this mode the whole
array must be clocked twice to achieve a single image. First, the whole array
is clocked line by line, in which a fast reset is applied to the line
followed immediately by a read of that line to determine the pixel
biases. Then the whole array is clocked line by line a second time at
the same speed as the first array clocking to read the integrated
pixel values. The efficiency is 50\% for the minimum integration
time.}
\item{{\bf Full MPIA mode}.  The problem with the previous mode is
that as soon as the array has been clocked the second time to read the
integrated pixel values, those pixels continue to integrate: they are
only reset once the first clocking (reset-read) of the next cycle is
started. Full MPIA improves upon this by having only a single 
type of pass through the whole array. The array is clocked line by
line starting with a read of the integrated pixels of a line, followed
by a fast reset of that line, and then a read of the pixel bias values
of that line.  These bias values are those which will be subtracted
from the integrated pixel values obtained the {\em next}\ time the
whole array is read. Once the whole array has been read, there is a
pause (if the integration time is above the minimum) and the cycle
repeated for as many images (``repeats'') as required. Essentially no
time is now wasted, and for any integration time the efficiency of
this mode is almost 100\%.}
\end{enumerate}
Whereas the first mode is a completely array-oriented approach, the
full MPIA mode is a completely line-oriented approach. DCR with fast
reset is essentially a mix of the two.  Full MPIA mode is more
efficient because each line is reset and the bias values read
immediately after reading the integrated pixel values: DCR mode with
fast reset waits until the whole array has been read before reseting.
To obtain just a single image, DCR with fast reset takes the same time
as full MPIA mode, but for a sequence of many repeats, full MPIA mode
is much quicker (Table~\ref{readouttable}).  The relative speeds for
different integration times of the three modes has been empirically
determined at MPIA using HAWAII arrays (Fig.~\ref{readouteff}).  Note
that the minimum integration time with full MPIA mode is twice that of
DCR mode with fast reset (0.25 rather than 0.125s for the HAWAII-2),
but this is not a problem for Omega 2000.

\begin{table}
\begin{center}
\caption{Comparison of double correlated readout modes.}
\label{readouttable}
\vspace*{1ex}
\begin{tabular}{lll}
\hline
	      			& Full MPIA mode	& DCR with fast reset	\\
        			& (read-reset-read)	& (reset-read.read)	\\
\hline
Minimum integration time   	&  0.262s     		&	0.131s		\\
Correlated image frequency      &  4Hz  		&	4Hz		\\
Single image efficiency   	& 50\%			&  	50\%		\\
Stack of 10 images efficiency   & 91\%  		&	50\%		\\
Time for 60s integration    	& 60.262s     		& 	90.00s		\\
(0.262s integration per frame)	& & \\
\hline
\end{tabular}
\end{center}
\end{table}

\subsection{Data Acquisition and Control Software}

The MPIAs GEneric InfraRed camera Software
(GEIRS)~\cite{bizenberger_98a} will also be used for Omega 2000. The
new challenge is the 26 times higher data rate compared to the
1K$\times$1K HAWAII detector on existing instruments.  The data
acquisition software must be able to sustain this data rate
continuously, apply necessary preprocessing tasks, have data
visualization control and save the data in time to non-volatile
storage disks. The data are acquired by either one or two 30 MHz
16bit parallel interfaces.

The handling of the increased data rate can only be organised in a
current state of the art symmetrical multiprocessor system (e.g.\ a
SPARC Ultraserver 450), where nearly the full memory bandwidth is
consumed for the basic tasks needed to control the camera system and
human interfacing. The software package is already optimised for
minimal usage of memory by sharing data memory between all processes.
To keep the jobs synchronised without loss of efficiency, further
parallelization making optimal use of the multiprocessing capabilities
has to be implemented. The granularity of the shared memory buffers
for the parallel processing also has to be reorganised to make more
efficient use of the increased internal working memory.  Although this
can be done with standard computer architecture, there remains the
option of using distributed processing on standard hardware with
distributed memory (the so-called ``multiple instructions on multiple
data'' organization).

\end{document}